\documentclass[aps,prc,twocolumn,showpacs,showkeys,superscriptaddress]{revtex4}
\usepackage{graphicx,amsmath,amssymb,bm}

\newcommand{\kf}{k_{\rm F}}
\newcommand{\as}{a_s}

\newcommand{\be}{\begin{equation}}
\newcommand{\ee}{\end{equation}}
\newcommand{\fm}{\, \text{fm}}

\newcommand{\mev}{\, \text{MeV}}
\newcommand{\gcq}{\, \text{g}/\text{cm}^{3}}

\newcommand{\im}{{\rm Im}}

\begin{document}

\title{The Neutrino Response of Low-Density Neutron Matter from the Virial
Expansion}
\author{C.J. Horowitz}
\email[E-mail:~]{horowit@indiana.edu}
\affiliation{Nuclear Theory Center and Department of Physics, 
Indiana University, Bloomington, IN 47408}
\author{A. Schwenk}
\email[E-mail:~]{schwenk@u.washington.edu}
\affiliation{Department of Physics, University of Washington,
Seattle, WA 98195-1560}

\begin{abstract}
We generalize our virial approach to study spin-polarized neutron matter and 
the consistent neutrino response at low densities. In the long-wavelength 
limit, the virial expansion makes model-independent predictions for the 
density and spin response, based only on nucleon-nucleon scattering data. 
Our results for the neutrino response provide constraints for 
random-phase approximation or other model calculations, and we compare
the virial vector and axial response to response functions used in supernova
simulations. The virial expansion is suitable to describe matter near the
supernova neutrinosphere, and this work extends the virial equation
of state to predict neutrino interactions in neutron matter.
\end{abstract}

\pacs{21.65.+f, 26.50.+x, 25.30.Pt, 97.60.Bw}
\keywords{Low-density neutron matter, neutral-current neutrino scattering,
virial expansion}

\maketitle 

\section{Introduction}

Neutrinos radiate $99 \%$ of the energy in core-collapse supernovae. 
The scattering of neutrinos and the physics of the explosion are most
sensitive to the properties of low-density nucleonic 
matter~\cite{Mezza,RPLP}, which is
a complex problem due to strong coupling with large scattering lengths,
clustering in nuclear matter and the non-central nature of nuclear
interactions. For low densities and high temperatures, the virial
expansion provides a tractable approach to strong interactions, and
in previous works we have presented the virial equation of state of 
low-density nucleonic matter~\cite{vEOSnuc,vEOSneut}. The predicted large
symmetry energy at low densities has been confirmed in near Fermi
energy heavy-ion collisions~\cite{sym}.

The virial approach can be used to describe matter in thermal 
equilibrium around the neutrinosphere in supernovae. The temperature
of the neutrinosphere is roughly $T \sim 4 \mev$ from about
$20$ neutrinos detected in SN1987a~\cite{sn1987a1,sn1987a2}, and
the density follows from known cross sections
of neutrinos with these energies $n \sim 10^{11}-10^{12} \gcq$.
For neutron matter, the virial expansion in terms of the fugacity $z = 
e^{\mu/T}$ is valid for
\be
n = \frac{2}{\lambda^3} \, z + {\mathcal O}(z^2) \, \lesssim \,
4 \cdot 10^{11} \, (T/\text{MeV})^{3/2} \gcq \,,
\ee
where we require $z < 1/2$ and $\lambda$ denotes the thermal wavelength 
$\lambda = (2\pi/mT)^{1/2}$. Therefore, the virial approach makes
model-independent predictions for the conditions of the
neutrinosphere, based only on the experimental scattering data.

In this paper, we use the virial expansion to describe how neutrinos
interact with low-density neutron matter. We focus on 
neutral-current interactions, and leave charged-current reactions 
and nuclear matter to future works. Our long-term goal is 
a reliable equation of state and consistent neutrino
response for supernovae.

The free cross section per particle for neutrino-neutron elastic 
scattering is given by~\cite{opacities}
\be
\frac{1}{N} \frac{d\sigma_0}{d\Omega} = \frac{G_{\rm F}^2 E_\nu^2}{4 \pi^2} 
\biggl(C_a^2\,(3-\cos\theta) + C_v^2 \, (1+\cos\theta) \biggr) \,,
\label{sigmanun}
\ee
where $G_{\rm F}$ is the Fermi coupling constant, $E_\nu$ the neutrino 
energy, and $\theta$ the scattering angle. The weak axial coupling is $C_a
=g_a/2$, with $g_a = 1.26$ the axial charge of the nucleon. The weak vector 
charge is $C_v = -1/2$ for scattering from a neutron. Eq.~(\ref{sigmanun}) 
neglects corrections of order $E_\nu/m$ from weak magnetism and other 
effects~\cite{weakmag}.

In the medium, this cross section is modified by the vector response 
$S_v(q)$ and the axial response $S_a(q)$
\be
\frac{1}{N} \frac{d\sigma}{d\Omega}=\frac{G_{\rm F}^2 E_\nu^2}{16 \pi^2} 
\biggl(g_a^2\,(3-\cos\theta) S_a(q) + (1+\cos\theta) S_v(q)\biggr) \,,
\label{dsigmamed}
\ee
where $S_v$ and $S_a$ describe the response of the system to density 
and spin fluctuations respectively, and $q = 2 E_\nu \sin(\theta/2)$ 
denotes the momentum transfer. We will discuss the approximations for 
Eq.~(\ref{dsigmamed}) in Sect.~\ref{nuresponse}. In the following,
we will use the virial expansion to provide model-independent results 
for the response in the long-wavelength ($q \rightarrow 0$) or
forward-scattering limit.

This paper is organized as follows. We extend the virial equation of
state to spin-polarized matter in Section~\ref{formalism} and derive
the consistent long-wavelength response. Further details on the
virial equation of state can be found in Refs.~\cite{vEOSnuc,vEOSneut}. 
In Section~\ref{results}, we present results for the spin virial
coefficients, the pressure and entropy of spin-polarized neutron
matter, and the neutrino response. We compare our results to
Brueckner calculations, and to random-phase approximation (RPA)
response functions. Finally, we conclude in Section~\ref{conclusions}.

\section{Formalism}
\label{formalism}

The virial expansion is a general, model-independent approach for 
a dilute gas, provided the fugacity is small and for temperatures
above any phase transitions. Under these conditions, the 
grand-canonical partition function can be expanded in powers of 
the fugacity. The second virial coefficient $b_2$ describes the 
$z^2$ term in this expansion and is directly related to the two-body 
scattering phase shifts~\cite{b2,huang}. The relation of the
third virial coefficient to three-body scattering is not
straightforward, and was only studied for special 
cases~\cite{b3Pais,b3Bedaque,b3Rupak}. The virial expansion is 
not a perturbative $\kf \as$ expansion, and its great advantage
is that it includes bound states and scattering resonances on an 
equal footing.

\subsection{Spin-Polarized Matter}

The virial equation of state is easily generalized to spin-asymmetric 
systems. For two spin components, we denote the chemical potential 
for spin up and spin down particles by $\mu_+$ and $\mu_-$, with 
fugacity $z_+ = e^{\mu_+/T}$ and $z_- = e^{\mu_-/T}$ respectively.
For the virial equation of state we expand the pressure in a power
series of the fugacities
\be
P=\frac{T}{\lambda^3} \bigl( z_+ + z_- 
+ b_{n,1} \, (z_+^2 + z_-^2) + 2 \, b_{n,0} \, z_+ z_- 
+ \mathcal{O}(z^3) \bigr) \,.
\label{p}
\ee
The second virial coefficients $b_{n,1}$ for like spins and $b_{n,0}$ 
for opposite spins are related to the two-particle partition function
and are given in terms of the scattering phase shifts in the next 
section. The densities follow from differentiating the pressure with 
respect to the fugacities. For the density of spin-up neutrons $n_+ 
= (\partial_{\mu_+} P)_T = z_+/T \, (\partial_{z_+} P)_T$ we thus have
\be
n_+ = \frac{1}{\lambda^3} \bigl( z_+ + 2 \, b_{n,1} \, z_+^2
+ 2 \, b_{n,0} \, z_+ z_- + \mathcal{O}(z^3) \bigr) \,,
\label{np}
\ee
and likewise for the density $n_-$ of spin-down neutrons
\be
n_- = \frac{1}{\lambda^3} \bigl( z_- + 2 \, b_{n,1} \, z_-^2
+ 2 \, b_{n,0} \, z_- z_+ + \mathcal{O}(z^3) \bigr) \,.
\label{nm}
\ee
The total density $n$ and the spin polarization $\Delta$ are then
given by
\be
n = n_+ + n_- \quad \text{and} \quad \Delta = \frac{n_+ -
n_-}{n_+ + n_-} \,.
\ee
In this work, we truncate the virial expansion after second order 
in the fugacities. This leads to an equation of state that is 
thermodynamically consistent.

The dependence of the total density and the spin polarization on
$z_+$ and $z_-$ can be inverted to yield the virial equation of 
state directly in terms of $P\bigl(z_+(n,\Delta,T),z_-(n,\Delta,T),
T\bigr)$. In practice, for a given spin polarization, we determine 
the spin-down fugacity as a function of the spin-up one $z_-(z_+,
\Delta,T)$, and generate the virial equation of state in tabular 
form for a range of $z_+$ values. This maintains
the thermodynamic consistency of the virial equation of state.

Finally, we will also discuss results for the entropy. The
entropy density $s=S/V$ follows from differentiating the pressure
with respect to the temperature $s=(\partial_T P)_{\mu_i}$. This 
leads to
\begin{align}
s &= \frac{5P}{2T} - n_+ \log z_+ - n_- \log z_- \nonumber \\[1mm]
&+ \frac{T}{\lambda^3} \bigl( b_{n,1}' \, (z_+^2 + z_-^2) + 2 \, 
b_{n,0}' \, z_+ z_- \bigr) \,,
\label{s}
\end{align}
where $b^\prime(T)=db(T)/dT$ denotes the temperature derivative of the 
virial coefficients.

\subsection{Spin Virial Coefficients}

The second virial coefficient $b_{n,1}$ describes the interaction
of two neutrons with the same spin projection. To this end, we 
generalize the second virial coefficient of the spin-symmetric
system~\cite{vEOSneut,b2,huang} to
\be
b_{n,1}(T)=\frac{2^{1/2}}{\pi T} \int_0^\infty dE \: e^{-E/2T} \, 
\delta^{\text{tot}}_1(E) - 2^{-5/2} \,,
\label{b1}
\ee
where $-2^{-5/2}$ is the free Fermi gas contribution and $\delta^{
\text{tot}}_1(E)$ is the sum of the isospin and spin-triplet elastic 
scattering phase shifts at laboratory energy $E$. This sum is over 
all partial waves with angular momentum $L$ and total angular momentum 
$J$ allowed by spin statistics, and includes a degeneracy factor 
$(2J+1)/(2S+1)$,
\begin{align}
\delta^{\text{tot}}_1(E) &= \sum_{L,J} \, \frac{2J+1}{3} \, 
\delta_{\,^{3}\text{L}_J}(E)
\nonumber \\[1mm]
&=\frac{1}{3} \, \delta_{^3\text{P}_0}+ \delta_{^3\text{P}_1}
+\frac{5}{3} \, \delta_{^3\text{P}_2}+ \ldots
\label{delta1tot}
\end{align}
The factor $1/(2S+1)=1/3$ arises because the same spin projection,
e.g., for up spins $M_S=+1$, is $1/3$ of the possibilities $M_S=-1,0,1$.
Note that we have neglected the effects of the mixing parameters due to 
the tensor force. We expect that their contributions are small for low
densities.

Two neutrons with opposite spin projections have a probability $1/2$ to 
be in spin $S=0$ or $S=1$ states, thus the second virial coefficient
for opposite spins $b_{n,0}$ is given by
\be
b_{n,0}(T)=\frac{2^{1/2}}{\pi T} \int_0^\infty dE \: e^{-E/2T} \, 
\delta^{\text{tot}}_0(E) \,,
\label{b0}
\ee
where $\delta^{\text{tot}}_0(E)$ is the sum of allowed 
isospin-triplet elastic scattering phase shifts with degeneracy 
factor $(2J+1)/(2(2S+1))$,
\begin{multline}
\delta^{\text{tot}}_0(E) = \sum_{S,L,J} \, \frac{2J+1}{2(2S+1)} \, 
\delta_{\,^{2S+1}\text{L}_J}(E) \\[1mm]
=\frac{1}{2}\,\delta_{^1\text{S}_0}+\frac{1}{6}\,\delta_{^3\text{P}_0}
+\frac{1}{2}\,\delta_{^3\text{P}_1}+\frac{5}{6}\,\delta_{^3\text{P}_2}
+\frac{5}{2}\,\delta_{^1\text{D}_2}+ \ldots
\label{delta0tot}
\end{multline}

The second virial coefficient for spin-symmetric neutron matter 
$b_n$ is the sum over like and opposite spins,
\be
b_n = b_{n,1} + b_{n,0} \,,
\ee
and consequently the sum of the total phase shifts given above
determines $b_n$ with
$\delta^{\text{tot}}(E)/2 = \delta^{\text{tot}}_0(E) +
\delta^{\text{tot}}_1(E)$~\footnote{The second virial coefficient
$b_n$ follows with $\delta^{\text{tot}}(E)/2$ instead of
$\delta^{\text{tot}}_1(E)$ in Eq.~(\ref{b1})~\cite{vEOSneut}.}.
In addition, we define the axial spin virial coefficient $b_a$
as
\be
b_a = b_{n,1} - b_{n,0} \,.
\ee
Thus, if only S-wave interactions are present, one has
\be
b_{n,1} = - 2^{-5/2}
\quad \text{and} \quad b_a = -b_n - 2^{-3/2} \,.
\ee

\subsection{Neutrino Response}
\label{nuresponse}

Neutrino scattering from a many-body system can be expressed in terms 
of the vector $S_v(q,w)$ and axial $S_a(q,w)$ dynamical response functions.
These describe the probability for a neutrino to transfer momentum $q$ 
and energy $w$ to the medium. Integrating over energy transfer, we define 
the static vector $S_v(q)$ and axial $S_a(q)$ response functions
\be
S_{v,a}(q) = \int_{-q}^q dw \, S_{v,a}(q,w) \,.
\ee
Here scattering kinematics limits the energy transfer to be space-like 
$|w| < q$. At low densities nucleons are nonrelativistic, and therefore 
we expect the vector response to have little strength in the time-like 
region so that
\be
S_v(q) \approx \int_{-\infty}^\infty dw \, S_v(q,w) \,.
\ee
The axial response can have contributions from multi-pair states in the 
time-like region even in the long-wavelength limit due to non-central nuclear 
interactions~\cite{multipair}. However, neutron matter at very low density can 
be described using a pion-less effective field theory where non-central 
interactions are sub-leading. Therefore, in this paper we approximate 
the axial response by
\be
S_a(q) \approx \int_{-\infty}^\infty dw \, S_a(q,w) \,.
\ee

The static structure factor for the density response is then given by
\begin{multline}
n \, S_v(q) = \frac{1}{Z} \, \sum_j e^{-\beta E_j} \int d^3{\bf r} \:
e^{i {\bf q} \cdot {\bf r}} \\[1mm]
\times \langle j | \psi^\dagger({\bf r})
\psi({\bf r}) \psi^\dagger(0) \psi(0) | j \rangle \,,
\end{multline}
where the sum is over all many-body eigenstates $| j \rangle$
with energy $E_j$, the partition function is $Z = \sum_j e^{-\beta E_j}$ 
and $\beta = 1/T$. For the spin response, the density operator is
replaced by the spin density $\psi^\dagger({\bf r}) {\bm \sigma}
\psi({\bf r})$.

In the long-wavelength limit, the vector response
of the spin-symmetric system is given by~\cite{LL} (see also
Appendix~B in~\cite{BS})
\be
S_v(q=0) = \frac{T}{(\partial P/\partial n)_T} \,.
\ee
For the symmetric system, the total chemical potential is $\mu =
(\mu_+ + \mu_-)/2$, with fugacity $z = \sqrt{z_+ z_-} = e^{\mu/T}$,
and the virial equation of state (see also Ref.~\cite{vEOSneut}) yields 
for the consistent vector response,
\be
S_v(q=0) = \frac {z}{n} \biggl( \frac{\partial n}{\partial z} \biggr)_T
= \frac{1+4 b_n z}{1+2 b_n z} \,.
\ee

Following Burrows and Sawyer~\cite{BS}, 
we define the spin-difference or axial chemical potential $\mu_a =
(\mu_+ - \mu_-)/2$ and the axial fugacity $z_a = \sqrt{z_+/z_-}$.
The axial response of the spin-symmetric system is then given by
\be
S_a(q=0) = \frac{z_a}{n} \frac{\partial}{\partial z_a} (n_+-n_-) 
\biggl|_{z_a=1} \,,
\ee
and the virial expansion, Eq.~(\ref{p}), leads to
\be
S_a(q=0) = 1 + \frac{2 b_a z}{1+2 b_n z} \,.
\ee
The long-wavelength limit of the axial response is also related to the 
spin susceptibility $\chi$,
\be
S_a(q=0) =
\frac{\chi}{\chi_{\rm F}} = \frac{n T}{(\partial^2 f/\partial \Delta^2
)_{n,T,\Delta=0}} \,,
\ee
where $f$ denotes the free energy density, and $\chi_{\rm F} = \mu_n^2 n/T$
is the spin susceptibility of a free neutron gas, with the neutron 
magnetic moment $\mu_n$. Finally, the response functions are normalized 
to unity in the low-density limit $S_v(0) = S_a(0) = 1$ for $z=n=0$.

\section{Results}
\label{results}

\subsection{Spin Virial Coefficients}

\begin{table}
\caption{The second virial coefficient $b_n$ and the axial virial
coefficient $b_a$ for different temperatures. The results labeled 
CIB take into account the effects due to charge-independence breaking 
(CIB) on the scattering length with $a_{nn} = -18.5 \fm$. 
We estimated an error of $< 5 \%$ for the higher temperatures 
$T \geqslant 25 \mev$ due to the truncation of the integration 
over the phase shifts at $E \leqslant 350 \mev$.}
\label{Table1}
\begin{ruledtabular}
\begin{tabular}{c|ccc|ccc}
$T [\text{MeV}]$ & $b_n$ & with CIB & $T \, b_n^\prime$ &
$b_a$ & with CIB & $T \, b_a^\prime$ \\[0.2mm] \hline
  1.00 &  0.288 &  0.251 &  0.032 & -0.641 & -0.604 & -0.031 \\[0.2mm]
  2.00 &  0.303 &  0.273 &  0.012 & -0.655 & -0.625 & -0.007 \\[0.2mm]
  3.00 &  0.306 &  0.279 &  0.004 & -0.655 & -0.629 &  0.006 \\[0.2mm]
  4.00 &  0.306 &  0.283 &  0.001 & -0.652 & -0.628 &  0.014 \\[0.2mm]
  5.00 &  0.306 &  0.285 &  0.000 & -0.648 & -0.627 &  0.020 \\[0.2mm]
  6.00 &  0.306 &  0.286 &  0.001 & -0.644 & -0.624 &  0.023 \\[0.2mm]
  7.00 &  0.307 &  0.288 &  0.002 & -0.640 & -0.621 &  0.026 \\[0.2mm]
  8.00 &  0.307 &  0.289 &  0.004 & -0.637 & -0.619 &  0.028 \\[0.2mm]
  9.00 &  0.308 &  0.291 &  0.007 & -0.634 & -0.616 &  0.029 \\[0.2mm]
 10.00 &  0.309 &  0.292 &  0.009 & -0.631 & -0.614 &  0.029 \\[0.2mm]
 12.00 &  0.310 &  0.295 &  0.013 & -0.625 & -0.610 &  0.029 \\[0.2mm]
 14.00 &  0.313 &  0.299 &  0.017 & -0.621 & -0.607 &  0.028 \\[0.2mm]
 16.00 &  0.315 &  0.302 &  0.020 & -0.617 & -0.604 &  0.026 \\[0.2mm]
 18.00 &  0.318 &  0.305 &  0.022 & -0.614 & -0.602 &  0.024 \\[0.2mm]
 20.00 &  0.320 &  0.308 &  0.023 & -0.612 & -0.600 &  0.021 \\[0.2mm]
 22.00 &  0.322 &  0.311 &  0.023 & -0.610 & -0.598 &  0.019 \\[0.2mm]
 24.00 &  0.324 &  0.313 &  0.022 & -0.608 & -0.597 &  0.018 \\[0.2mm]
 26.00 &  0.326 &  0.315 &  0.021 & -0.607 & -0.596 &  0.017 \\[0.2mm]
 28.00 &  0.327 &  0.317 &  0.018 & -0.606 & -0.595 &  0.016 \\[0.2mm]
 30.00 &  0.329 &  0.318 &  0.015 & -0.605 & -0.595 &  0.016 \\[0.2mm]
 35.00 &  0.330 &  0.321 &  0.004 & -0.602 & -0.593 &  0.018 \\[0.2mm]
 40.00 &  0.330 &  0.321 & -0.009 & -0.599 & -0.591 &  0.024 \\[0.2mm]
 45.00 &  0.328 &  0.319 & -0.025 & -0.596 & -0.588 &  0.031 \\[0.2mm]
 50.00 &  0.324 &  0.316 & -0.041 & -0.592 & -0.584 &  0.041 \\
\end{tabular} 
\end{ruledtabular}
\end{table}

We first calculate the virial coefficients $b_n$ and $b_a$
from the $T=1$ $np$ phase shifts obtained
from the Nijmegen partial wave analysis~\cite{nnphases}. This
neglects the small charge dependences in nuclear interactions. We have
included all partial waves with $L \leqslant 6$. For the higher
temperatures, $T \geqslant 25 \mev$, there is a $< 5\%$ error 
due to the truncation of the integration over the phase shifts at
 $E \leqslant 350 \mev$ (the extent of the partial wave analysis). 
This error was estimated by assuming constant total phase shifts and
varying the energy cutoff to $E > 350 \mev$.

Our results for the virial coefficients and their temperature
derivatives $T b^\prime(T)$ are listed in Table~\ref{Table1}.
As discussed in Ref.~\cite{vEOSneut}, the virial coefficients are
dominated by the large S-wave scattering length physics ($a_{np}
=-23.768 \fm$ and $a_{nn} = -18.5 \fm$), but effective range 
and higher partial wave contributions are noticeable. For
example, for $T = 5\mev$, the virial coefficients obtained
only from $a_{np}$ are $b_n = 0.44$~\cite{vEOSneut} and 
$b_a = -b_n - 2^{-3/2}=-0.80$. In the unitary limit where the
scattering length $|a_s| = \pm \infty$ and $\delta(E) = \pi/2$,
the second virial coefficients are independent of the temperature 
and given by $b_n = 3/2^{5/2} = 0.53$~\cite{Ho1} and $b_a = -
5/2^{5/2} = -0.88$. Therefore,
the virial expansion is well defined for resonant interactions,
in contrast to the $k_{\rm F} a_s$ expansion.

\begin{figure}[t]
\begin{center}
\includegraphics[scale=0.36,clip=]{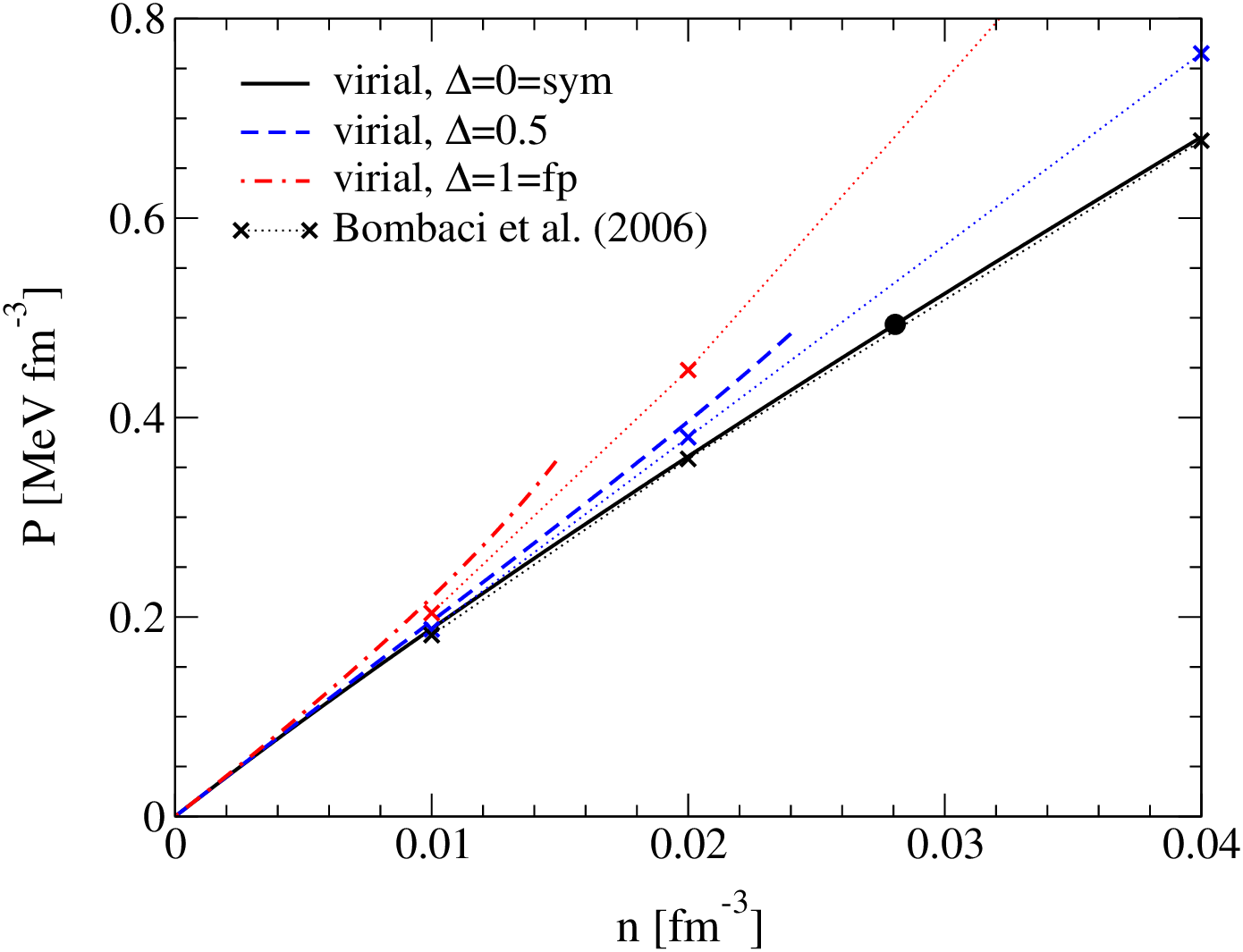}
\caption{(Color online) The pressure $P$ versus density $n$ 
for $T=20 \mev$ and various spin polarizations $\Delta = 0$ 
(symmetric), $0.5$ and $1$ (fully polarized). We also compare 
our results to Brueckner calculations of Bombaci 
{\it et al.}~(crosses with dotted lines)~\cite{Bombaci}
for $\Delta = 1, 0.5, 0$ (top to bottom). 
The circle indicates where the fugacity 
is $z =0.5$ for $\Delta=0$, and for the other spin polarizations
the virial curves end at $z=1$.}
\label{Fig1}
\vspace*{-4mm}
\end{center}
\end{figure}

We find that the second virial coefficients are approximately 
independent of temperature over a wide range, and consequently 
$T b^\prime(T) \approx 0$. As a result, the thermodynamic
properties of spin-polarized neutron matter and the long-wavelength 
response scale as a function of the fugacities, which depend
on density and temperature through $z_i(n_+/T^{3/2},n_-/T^{3/2})$
for $i=+$ and $i=-$. This
scaling can also be expressed in terms of the Fermi temperatures
$T_{{\rm F},i} \sim n_i^{2/3}$, and thus the properties of neutron
matter scale with $T/T_{{\rm F},i} \sim T/n_i^{2/3}$ only. 
In Ref.~\cite{vEOSneut}
we found that spin-symmetric neutron matter scales to a very good
approximation. The virial scaling symmetry is exact for cold atomic
gases tuned to a Feshbach resonance~\cite{Ho2} and has been
verified experimentally by Thomas {\it et al.}~\cite{Thomas}.

In Table~\ref{Table1}, we also study
the effects of charge-independence breaking (CIB) 
on the scattering length. We estimate CIB effects as discussed
in Ref.~\cite{vEOSneut}. CIB for the virial coefficients is largest 
for $T < 5 \mev$ and leads to a $10 \%$ reduction in magnitude of 
the virial coefficients.

\subsection{Pressure and Entropy of Spin-Polarized Matter}

\begin{figure}[t]
\begin{center}
\includegraphics[scale=0.36,clip=]{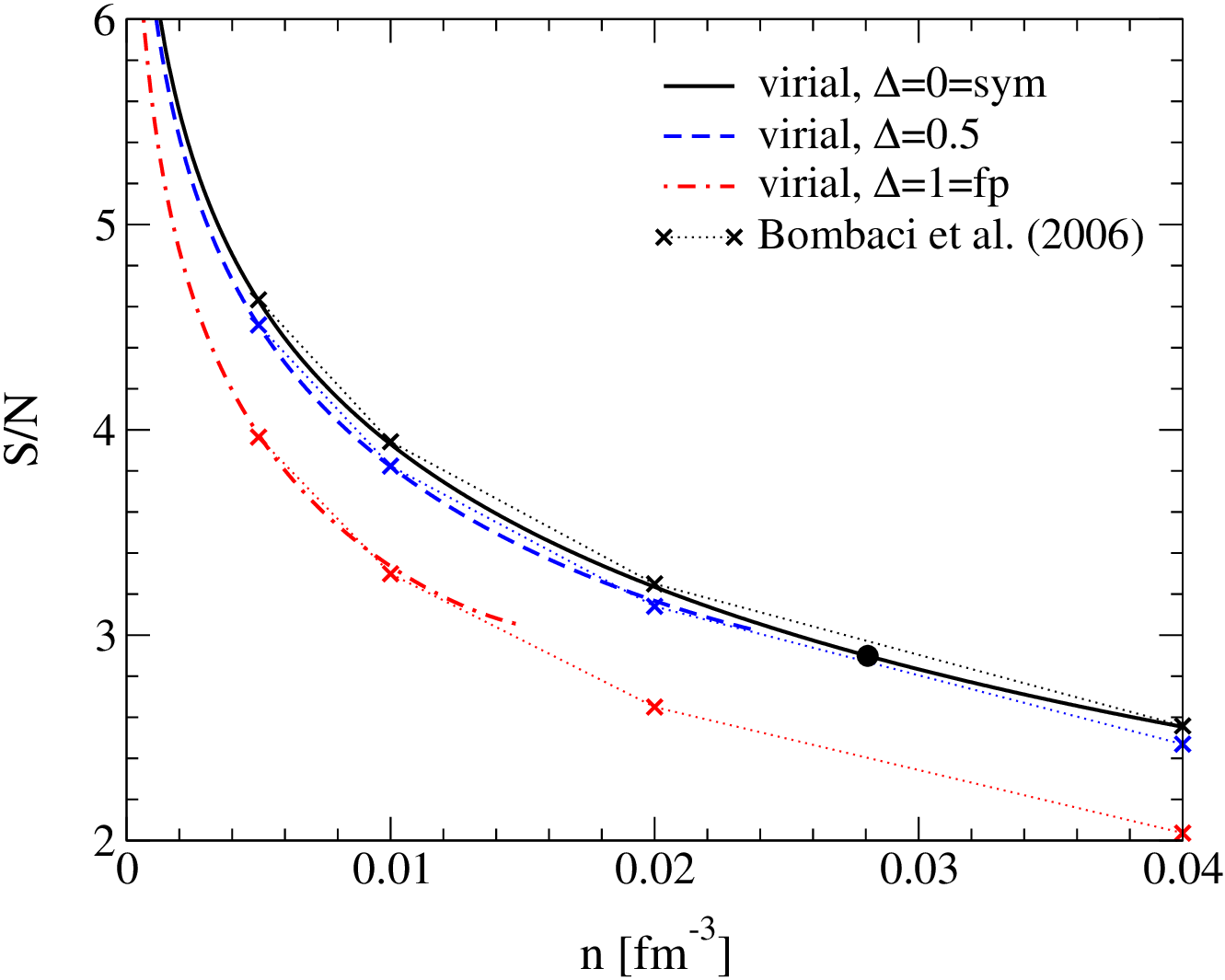}
\caption{(Color online) The entropy per particle $S/N$ 
versus density $n$ for $T=20 \mev$ 
and various spin polarizations $\Delta = 0$ 
(symmetric), $0.5$ and $1$ (fully polarized). We also compare 
our results to Brueckner calculations of Bombaci 
{\it et al.}~(crosses with dotted lines)~\cite{Bombaci}
for $\Delta = 0, 0.5, 1$ (top to bottom).
The circle indicates where the fugacity 
is $z =0.5$ for $\Delta=0$, and for the other spin polarizations
the virial curves end at $z=1$.}
\label{Fig2}
\vspace*{-4mm}
\end{center}
\end{figure}

We have previously found~\cite{vEOSneut} good agreement of the virial 
equation of state for spin-symmetric neutron matter with microscopic 
Fermi hyper-netted chain (FHNC) calculations of Friedman and
Pandharipande~\cite{FP} for densities up to $n \lesssim n_0/10$,
where $n_0 = 0.16 \, \text{fm}^{-3}$ is the saturation density
of symmetric nuclear matter, and published temperatures $T 
\geqslant 10 \mev$. For nuclear matter, the FHNC results fail to
describe clustering with alpha particles at low densities~\cite{vEOSnuc}.

Our virial results for the pressure and entropy of spin-polarized
neutron matter are shown in Figs.~\ref{Fig1} and~\ref{Fig2} for
$T=20 \mev$ and polarizations $\Delta = 0$ 
(symmetric), $0.5$ and $1$ (fully polarized). For this
temperature, we can compare the virial results to Brueckner
calculations of Bombaci {\it et al.}~\cite{Bombaci}. As shown
in Fig.~\ref{Fig2}, the Brueckner entropy agrees well with the
virial results.
For the pressure, the effects of a spin polarization are smaller, and
in addition there is some uncertainty in the Brueckner calculations,
since the pressure was obtained from the energy by a numerical
derivative. Before we discuss the neutrino response, we
note that it is difficult to calculate the long-wavelength
response at low densities from the Brueckner or FHNC results, since 
the response is obtained by differentiating the pressure.

\subsection{Neutrino Response}

\begin{figure}[t]
\begin{center}
\includegraphics[scale=0.36,clip=]{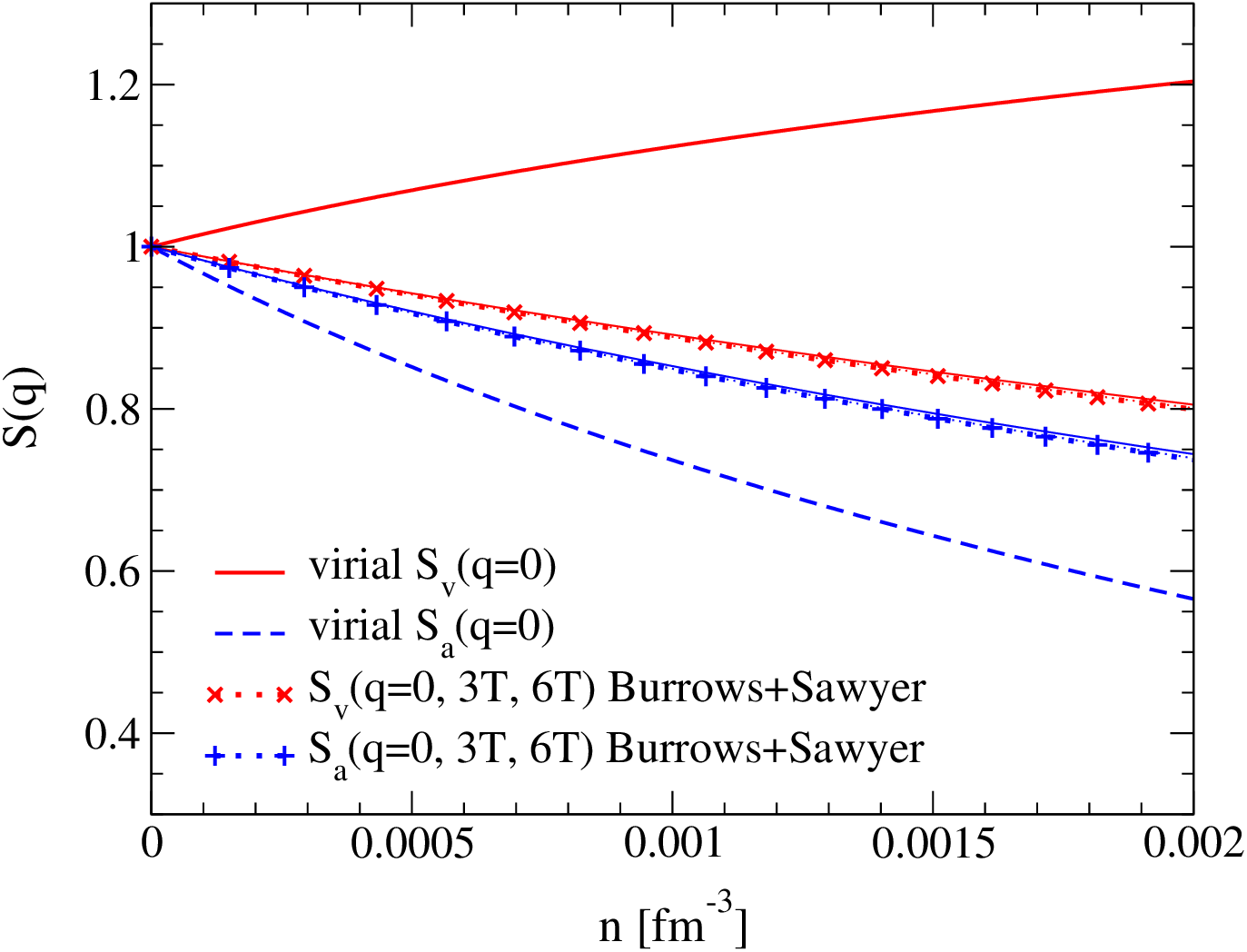}
\caption{(Color online) The vector and axial response of neutron
matter for $T = 4 \mev$. In addition to the long-wavelength
virial response, we also show the RPA response of Burrows and
Sawyer~\cite{BS} for neutron matter and various momentum
transfers $q=0, 3T$ and $6T$. For this density range, the fugacity in
the virial expansion is $z < 0.42$.}
\label{Fig3}
\vspace*{-4mm}
\end{center}
\end{figure}

Our virial results for the long-wavelength vector and axial response 
are presented in Fig.~\ref{Fig3}. The neutron-neutron interaction
is attractive at long distances and thus increases the probability 
to find two neutrons close together compared to a free neutron
gas. These density fluctuations increase the local weak charge and 
produce a vector response $S_v > 1$ for low-momentum transfers.
This is easily seen by expanding the vector response to 
lowest order in the density. With $z \approx n \lambda^3/2$, we
have
\be
S_v(q=0) \approx 1 + b_n n \lambda^3 > 1 \,,
\ee
since $b_n = 0.31$ from Table~\ref{Table1}. In a Landau-Fermi
liquid, the vector response is given by $S_v(0) = 1/(1+F_0) > 1$
for neutron matter, where the Landau parameter 
for the density-density interaction is $F_0 < 0$~\cite{RGnm}.

In contrast, the spin-spin interaction is repulsive (this follows 
from the Pauli principle, when the density-density interaction is
attractive), and the virial axial response gives $S_a < 1$
for low-momentum transfers. This is seen in the low-density
limit,
\be
S_a(q=0) \approx 1 + b_a n \lambda^3 < 1 \,,
\ee
where $b_a = -0.65$ from Table~\ref{Table1}. Analogous to the
vector response, the axial response for a Landau-Fermi liquid
is given by $S_a(0) = 1/(1+G_0) < 1$ for neutron matter, since 
the Landau parameter for the spin-spin
interaction is $G_0 > 0$~\cite{RGnm}. Although the virial densities
and temperatures are not in a Fermi liquid regime
($z \sim (T_{\rm F}/T)^{3/2} \ll 1$), the deviation of the vector
and axial response from a free gas is determined by nuclear 
interactions, and thus is the same for low and high temperatures. 

\subsection{Comparison to RPA calculations}

Most present calculations of the neutrino response are based on 
the random-phase approximation (RPA)~\cite{BS,RPLP,HG}, 
which gives the linear 
response of a mean-field ground state to neutrinos. The RPA
response thus neglects clustering and is incorrect for nuclear 
matter at subnuclear densities~\cite{response}. Since there
is no clustering in neutron matter, a comparison
of the virial with RPA response assesses the interactions
used in present RPA calculations, as well as the random-phase
many-body approximation for low densities and high temperatures.

\begin{figure}[t]
\begin{center}
\includegraphics[scale=0.36,clip=]{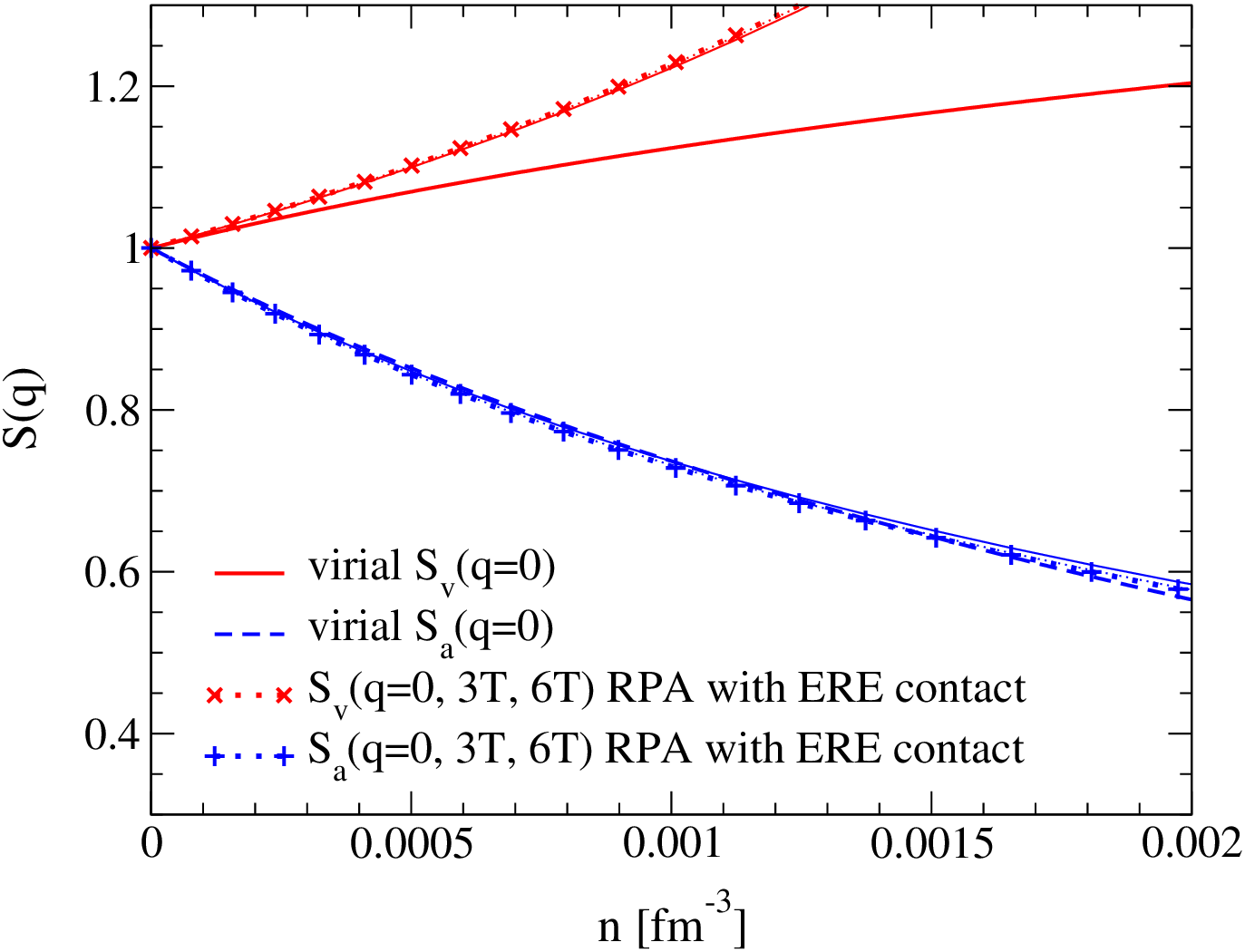}
\caption{(Color online) The long-wavelength virial response of
neutron matter for $T = 4 \mev$ is compared to the RPA response 
with effective-range expansion (ERE) contact interactions and 
average neutron energy $\langle E \rangle = 3T/2$.}
\label{Fig4}
\vspace*{-4mm}
\end{center}
\end{figure}

As an example, we compare our virial results to the nonrelativistic
RPA calculations of Burrows 
and Sawyer~\cite{BS}, where the RPA interaction is chosen to 
reproduce Landau-Fermi liquid parameters for symmetric nuclear
matter. We compare to the approach of Ref.~\cite{BS}, because these
results have been used in supernova simulations~\cite{Buras} and 
they are somewhat simpler and thus more transparent
than Refs.~\cite{RPLP,HG}.
Since Burrows and Sawyer do not present results for pure
neutron matter, we have calculated the RPA response following
Ref.~\cite{BS}. For completeness, we give the necessary equations
in Appendix~\ref{RPA}. For low-density neutron matter, the
effective mass is well approximated by the free mass~\cite{RGnm},
and we thus use $m^*/m=1$.

In Fig.~\ref{Fig3}, we compare the RPA results for $T = 4 \mev$
to the virial response. We find that the RPA axial response is 
repulsive ($S_a < 1$) and on a qualitative level similar to the virial 
response. However, the RPA vector response is also repulsive, in contrast 
to our virial result. This is because Ref.~\cite{BS}
uses Landau parameters of symmetric nuclear matter for all
proton fractions. In particular, 
Burrows and Sawyer use for the spin-independent
part of the interaction $F_0 + F_0' \, {\bm \tau}_1 \cdot {\bm \tau}_2$, 
with $F_0 = -0.28$ and $F_0' = 0.95$~\cite{BS}, and
the density-density interaction for neutrons is thus $F_0(T=1) =
F_0 + F_0' =0.67$. This makes the incorrect assumption that 
induced interactions in nuclear and neutron matter 
are identical. For the virial coefficient $b_n$,
the total phase shift is attractive~\cite{vEOSneut}. This leads
to an attractive vector response ($S_v > 1$) at low densities 
and low-momentum transfers. The RPA results of Refs.~\cite{RPLP,HG}
have an attractive vector response, consistent with the
mean-field equation of state. However, the axial
interaction of Ref.~\cite{HG} is not constrained at the mean-field
level and may be more poorly determined.

\begin{figure}[t]
\begin{center}
\includegraphics[scale=0.36,clip=]{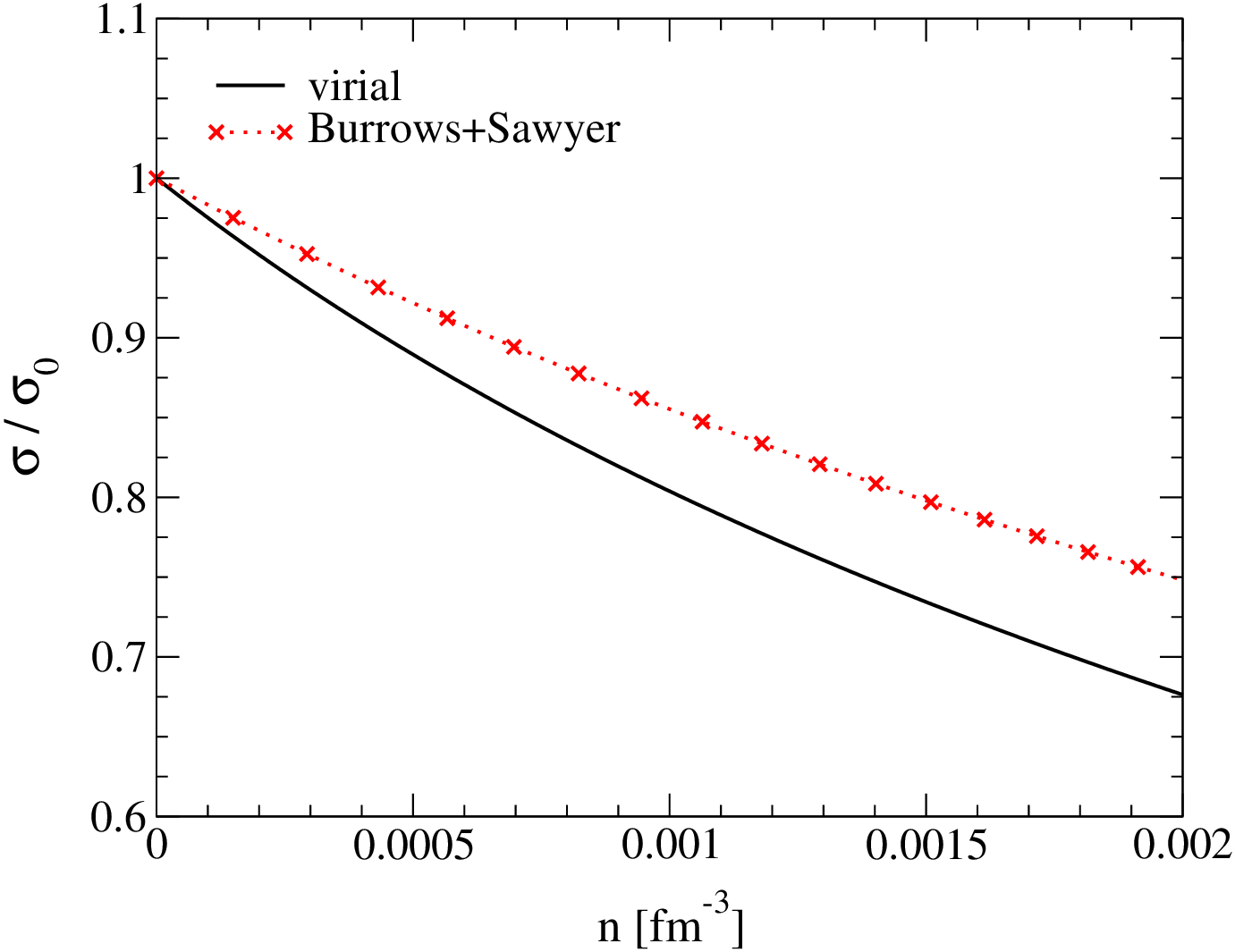}
\vspace*{-2mm}
\caption{(Color online) The total response of neutron matter given 
by the ratio of the total cross section for elastic neutrino-neutron 
scattering in the medium compared to free space. The results shown are 
for $T = 4 \mev$ and neglect the small momentum dependence of the
vector and axial response.}
\label{Fig5}
\vspace*{-4mm}
\end{center}
\end{figure}

The RPA provides a model
to study the momentum dependence of the response functions.
For a neutrino with energy $E_\nu = 3T$, the maximum momentum
transfer is $q_{\rm max} = 2 E_\nu = 6T$. In addition to
the long-wavelength response, Fig~\ref{Fig3} shows the RPA 
results for various momentum transfers.
This demonstrates that the RPA response has a very weak momentum 
dependence. Consequently, the long-wavelength response 
provides strong constraints for all relevant momentum transfers.

Next, we calculate the RPA response, when we use contact 
interactions that are constrained by nucleon-nucleon scattering. 
In order to obtain cutoff-independent results and correctly 
include the large scattering length and effective range at 
low density, it is necessary to sum particle-particle ladders 
and work with the $T$ matrix (see also~\cite{dEFT}). This leads to 
Landau parameters $f_0 + g_0 \, {\bm \sigma}_1 
\cdot {\bm \sigma}_2$ ($F_0 = m \, k_{\rm F} f_0/\pi^2$ and
$G_0 = m \, k_{\rm F} g_0/\pi^2$) for the antisymmetrized 
interaction with 
\be
f_0 = \frac{2 \pi/m}{1/a_{np} - m \, r_e \, E/2} \quad \text{and} \quad
g_0 = - f_0 \,,
\ee
where $r_e = 2.68 \fm$ is the effective 
range and $E$ denotes the relative energy. In order to make a simple
estimate, we take an average relative energy $\langle E \rangle = 
\langle (p_1 - p_2)^2 \rangle/4m = 3T/2$ and 
calculate the RPA response with these Landau parameters. The
resulting vector and axial response is shown in Fig.~\ref{Fig4}.
The axial response agrees nicely with our virial result, but for 
the vector response there is only a good agreement at low 
densities. The differences at higher densities could be
due to using an average energy, since the vector response
is more sensitive to the latter.

Finally, we show the total response of neutron matter for $T = 4
\mev$ in Fig.~\ref{Fig5}. The total response is given by the ratio
of the total cross section for elastic neutrino-neutron scattering 
in the medium compared to free space. We neglect the small momentum 
dependence of the vector and axial response, and thus have $\sigma /
\sigma_0 = (6 g_a^2 S_a(0) + 2 S_v(0)) / (6 g_a^2 + 2)$. We find
for example a factor $0.72$ reduction of the total response at
$n = 0.0016 \, \text{fm}^{-3} = n_0/100$. This is $10 \%$ 
larger compared to the RPA response of Burrows and Sawyer.

\section{Conclusions}
\label{conclusions}

We have extended our virial approach to study spin-polarized 
neutron matter and the consistent long-wavelength response.
The virial expansion is suitable to describe matter near the
supernova neutrinosphere, and this work extends the virial 
equation of state~\cite{vEOSnuc,vEOSneut} to predict neutrino 
interactions in neutron matter. Our results include the 
physics of the large neutron-neutron scattering length in a tractable 
way. We have found that the spin virial coefficients are
approximately temperature independent over a wide range.
The properties of spin-polarized neutron matter and the 
response therefore scale with density and temperature as 
discussed in Ref.~\cite{vEOSneut}.

The virial expansion was used to make model-independent 
predictions for the pressure and entropy of spin-polarized
matter, for the vector and axial response, and the cross 
section for neutrino-neutron scattering in the medium.
The virial pressure and entropy of spin-polarized neutron 
matter are similar to Brueckner results~\cite{Bombaci}, but the 
virial approach has a well-defined range of validity and is 
directly based on scattering data.

The virial equation of state predicts an attractive vector 
and a repulsive axial response in the long-wavelength
limit at low densities. The total neutrino response is
suppressed in matter compared to the response of a free
neutron gas. This provides a benchmark for many-body 
calculations of the response functions. As an
example, our results for the neutrino response disagree 
with the RPA response of Burrows and Sawyer~\cite{BS}
due to the interaction model used for the latter. The RPA 
was used to study the momentum dependence of the 
response functions. We have found a very weak dependence
on momentum transfer (independent of sign and magnitude
of the interaction). We therefore conclude that the 
long-wavelength virial response provides strong constraints 
at low densities for all relevant momentum transfers.

Important areas of future work are the extension
of these techniques to charged-current interactions
and to the neutrino response in nuclear matter~\cite{response}.
In addition, a generalization of the virial expansion
beyond Section~\ref{nuresponse} may offer insights to the 
effects of multi-pair states on the long-wavelength response 
at low densities~\cite{multipair}. The third virial coefficient
can be used to provide error estimates~\cite{vEOSnuc,vEOSneut}. 
For the neutrino response, an effective field theory calculation 
of the dominant large scattering length contributions to the 
third spin virial coefficients would be very useful.

\vspace*{-5mm}
\section*{Acknowledgments}

We thank Isaac Vidana for providing us with the Brueckner results. 
This work is supported by the US Department 
of Energy under Grants No.~DE--FG02--87ER40365 and DE--FG02--97ER41014.

\vspace*{5mm}
\appendix

\section{RPA response}
\label{RPA}

The static structure function $S(q)$ is given in terms of the
polarization function $\chi(q,\omega)$ by
\be
S(q) = \frac{1}{\pi} \int d\omega \: \frac{\im\,  \chi(q,\omega)}{1-
e^{-\omega/T}} \,,
\ee
where the polarization function in RPA reads
\be
\chi(q,\omega) = \frac{\Pi^0(q,\omega)}{1 - v_0 \, \Pi^0(q,\omega)} \,.
\ee
The Landau interaction used by Burrows and Sawyer~\cite{BS} is
$v_0 = f_0 = 1.76 \cdot 10^{-5} \, \text{MeV}^{-2}$ for the
density response and $v_0 = g_0 = 4.50 \cdot 10^{-5} \, 
\text{MeV}^{-2}$ for the spin response.
The real and imaginary parts of the free
polarization $\Pi^0(q,\omega)$ are derived in the Appendix 
of Ref.~\cite{BS},
\begin{align}
{\rm Re} \, \Pi^0(q,\omega) &= \frac{m^2}{2 \pi^2 q \beta} \,
\int\limits_0^\infty \, \frac{ds}{s} \ln \biggl[
\frac{1 + e^{\beta \mu -(s+Q)^2}}{1 +
e^{\beta \mu -(s-Q)^2}} \biggr] \nonumber \\
&+ \omega \to - \omega \,,
\label{repi} \\
\im \, \Pi^0(q,\omega) &= \frac{m^2}{2 \pi q \beta} \,
\ln \biggl[ \frac{1 + e^{\beta \mu -Q^2}}{1 +
e^{\beta (\mu-\omega) -Q^2}} \biggr] \,,
\label{impi}
\end{align}
with $Q=\sqrt{m\beta/2} \, ( -\omega/q + q/(2m))$. Finally, the
density is given by
\be
n = 2 \int \frac{d^3p}{(2\pi)^3} \frac{1}{1+ e^{\beta (p^2/(2m)
-\mu)}} \,,
\ee
which determines the chemical potential for Eqs.~(\ref{repi},\ref{impi}).

\end{document}